\documentclass[10pt,letterpaper,twocolumn]{article}

\usepackage{ol2}
\usepackage[draft]{hyperref}
\usepackage{amsmath}

\renewenvironment{abstract}
{
\noindent \begin{center}
{\footnotesize Compiled \today}
\vskip4pt \begin{minipage}{34.25pc} \parindent.2in \noindent \footnotesize \rm
}
{
\\ \hfil \end{minipage} \end{center}
}

\begin{document}

\twocolumn[

\title{Broadband transmission properties of multilayered structures}

\author{Victor Grigoriev$^*$ and Fabio Biancalana}

\address{
Max Planck Institute for the Science of Light, G\"{u}nther-Scharowsky-Str. 1, Bau 26, Erlangen 91058, Germany
$^*$Corresponding author: victor.grigoriev@mpl.mpg.de
}

\begin{abstract}
The formalism of the scattering matrix is applied to describe the transmission properties of multilayered structures with deep variations of the refractive index and arbitrary arrangements of the layers. We show that there is an exact analytical formula for the transmission spectrum, which is valid for the full spectral range and which contains only a limited number of parameters for structures satisfying the quarter-wave condition. These parameters are related to the poles of the scattering matrix, and we present an efficient algorithm to find them, which is based on considering the ray propagation inside the structure and subsequent application of the harmonic inversion technique. These results are significant to analyze the reshaping of ultrashort pulses in multilayered structures.
\end{abstract}


]

\noindent
Multilayered structures with periodic arrangement of the layers represent the simplest example of photonic crystals. Such structures are widely used as distributed reflectors, spectral filters and can be applied for compression or reshaping of ultrashort pulses \cite{Gaponenko2010}. A lot of attention has been paid to the question of how to modify the properties of multilayered structures by introducing artificial defects. As a more general case, structures with quasiperiodic and deterministically aperiodic arrangements of the layers were considered \cite{Albuquerque2003}. The standard technique to compute the transmission spectrum of multilayered structures is based on the transfer matrix method. However, being a strictly numerical method, it does not provide a proper understanding of the transmission properties from the physical point of view.

In this paper, we show that there is an exact analytical formula for the transmission spectrum which is applicable for multilayered structures with a deep variation of the refractive index and arbitrary arrangement of the layers. It is valid for the full spectral range and contains only a limited number of parameters for structures satisfying the quarter-wave condition. The knowledge of the precise analytical formula for the broad band transmission allows one not only to compute easily many important characteristics such as the group velocity dispersion or the photonic density of modes, but also to analyze the propagation of arbitrary signals through the structure in the time domain.

Any multilayered structure can be considered as a black box, the input and output from which are related by the scattering matrix. It is known that the poles and zeros of the scattering matrix determine the properties of a system uniquely \cite{Chong2010}. Taking into account that the components of the scattering matrix can be interpreted as reflection and transmission coefficients, one can describe the transmission spectrum $T(\omega)$ of an arbitrary multilayered structure by the following formula
\begin{equation}
\label{eqTransmInf}
T(\omega) = \sum \limits_{p = - \infty}^{\infty}
{\frac{\sigma_p}{\omega - \omega_p}},
\end{equation}
where $\sigma_p$ is the strength of the resonance at $\omega_p$, and the sum is taken over all resonances. The resonances are associated with poles in the transmission spectrum $T^{-1}(\omega_p) = 0$. For media without gain all poles are located in the lower part of the complex frequency plane ${\rm{Im}}(\omega_p) < 0$. The strength of the resonances can be found as $\sigma_p = [{\rm{d}} T^{-1}(\omega) / {\rm{d}} \omega]_{\omega  = \omega_p }^{-1}$.

\begin{figure*}
\centerline{\includegraphics[width=170mm]{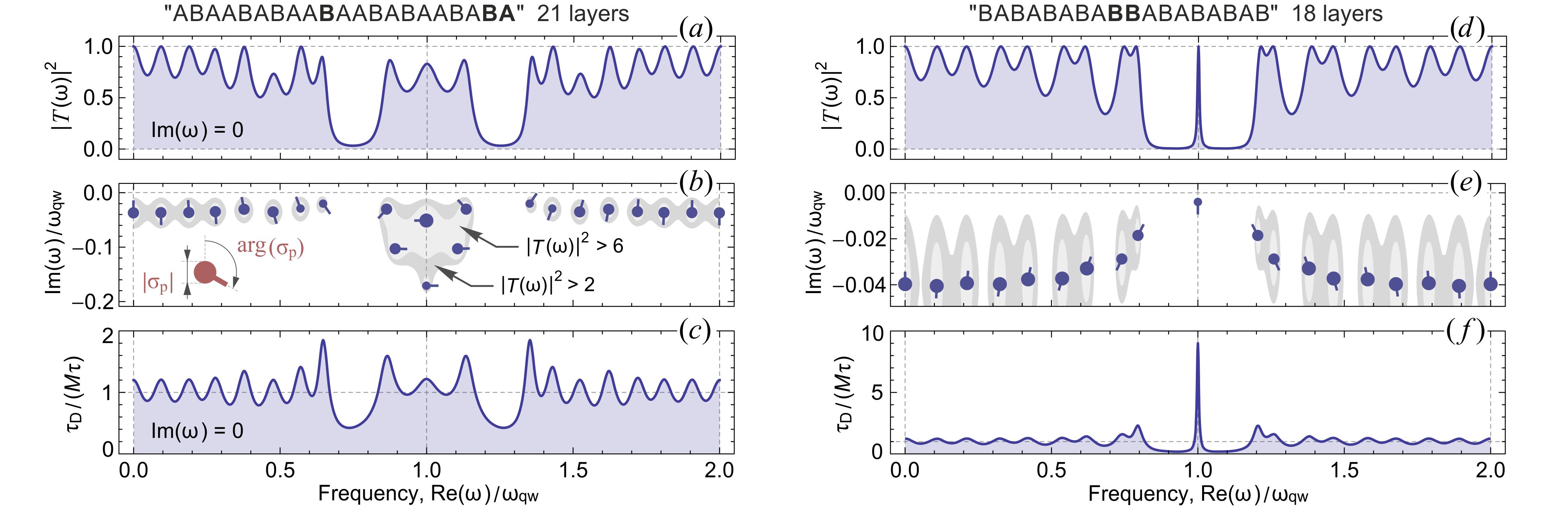}}
\caption{\fontsize{9}{10}\selectfont{\label{fig1} (Color online)
(a,~d)~Transmission spectrum of a quarter-wave structure described by a binary sequence. The letters "A" and "B" correspond to layers with the refractive indices 1.55 and 2.3, respectively.
(b,~e)~Positions of the poles in the complex frequency plane. In contrast to (a,~d), where ${\rm Im}(\omega) = 0$ and $|T(\omega)| \leq 1$, transmission goes to infinity in the vicinity of poles, and the gray areas display regions where it exceeds a few fixed values.
(c,~f)~Group delay as a function of frequency.
}}
\end{figure*}

In general, the number of resonances is infinite, but due to the time-reversal symmetry $
T(- \omega^*) = T^*(\omega)$, which means that the poles exist in pairs $\{ \omega_p , - \omega_p^* \}$ and that their strengths are related as $\sigma(- \omega_p^*) = - \sigma^*(\omega_p)$. Those multilayered structures, which satisfy the quarter-wave condition $n_m d_m = \lambda_{\rm{qw}} / 4$ at the wavelength $\lambda_{\rm{qw}}$ for each layer $m$ with the refractive index $n_m$ and thickness $d_m$, have the additional symmetry $T(\omega  + 4 k \omega_{\rm{qw}}) = T(\omega)$, where $k$ is an arbitrary integer, $\omega_{\rm{qw}}$ is the frequency at which the quarter-wave condition holds ($\omega_{\rm{qw}} / c = 2 \pi / \lambda_{\rm{qw}}$). Therefore, there is only a limited number of independent poles $P$ in the interval $(-2 \omega_{\rm{qw}}, 2 \omega_{\rm{qw}}]$, and the positions of all the others can be found by adding $4k \omega_{\rm{qw}}$.

The inverse Fourier transform of Eq.~(\ref{eqTransmInf}) gives the response function $G(t)$ of the structure to the excitation in the form of the Dirac delta function $\delta(t)$
\begin{equation}
\label{eqGtTemp}
G(t) =  - i h(t)
    \sum\limits_{p = 1}^P {\sigma_p {\rm{e}}^{- i \omega_p t}}
    \sum\limits_{k = - \infty }^{\infty} {\rm{e}}^{- i 4 k \omega_{\rm{qw}} t},
\end{equation}
where $h(t)$ is the Heaviside step function, and the periodicity of the poles for the quarter-wave structures was used explicitly. Applying the following property of the Dirac comb $
\sum\nolimits_{n = - \infty}^{\infty} {\delta (t - n \tau)} =
\sum\nolimits_{k = - \infty}^{\infty} {\tau^{-1} \exp [- i k (2 \pi / \tau) t ]}
$ leads to
\begin{equation}
\label{eqGt}
G(t) = \sum\limits_{n = 0}^{\infty}  {a_n \delta (t - n \tau)},
\end{equation}
where $\tau = \pi / (2 \omega_{\rm{qw}}) = n_m d_m / c$ has the meaning of time necessary for a signal to go through a quarter-wave layer. The coefficients in Eq.~(\ref{eqGt}) are defined as $a_n = - \sum\nolimits_{p = 1}^P {i \sigma_p \tau \exp(- i n \omega_p \tau)}$, and thus the response function can be viewed as a sum of exponentially decaying modes, which are sampled over the discrete intervals $\tau$. It is worth noting that all coefficients $a_n$ are real because $G(t)$ is real by definition. Moreover, the signal at the output cannot appear immediately, and $a_n  = 0$ for  $n < M$, where $M$ is the total number of layers in the structure. These causality relations emphasize that $\sigma_p$ are not independent parameters and are related to $\omega_p$.

The Fourier transform of Eq.~(\ref{eqGt}) gives
\begin{equation}
\label{eqTransmTemp}
T(\omega) = \sum\limits_{n = 0}^\infty {a_n {\rm{e}}^{i n \omega \tau}} = -
\sum\limits_{p = 1}^P {i \sigma_p \tau
\sum\limits_{n = 0}^\infty {{\rm{e}}^{i n(\omega - \omega_p)\tau}}},
\end{equation}
which shows that $a_n$ are equal to the Fourier components of the transmission spectrum. Taking the sum of the geometric progression in Eq.~(\ref{eqTransmTemp}) leads to
\begin{equation}
\label{eqTransmExp}
T(\omega) = \sum\limits_{p = 1}^P {
\frac{i \sigma_p \tau} {\exp [i (\omega - \omega_p) \tau] - 1}
}.
\end{equation}
Therefore, the transmission through the quarter-wave structures can be described by a formula which does not involve infinite series and has only a limited number of parameters. We checked that it gives exactly the same results as the transfer matrix method over the full spectral range. Two examples are shown in Fig.~\ref{fig1}, which correspond to multilayered structures based on the Fibonacci sequence of the 7th order (on the left) and a periodic sequence with a defect (on the right).

The periodicity of the poles could be used directly to obtain a few other formulas for the transmission. The repeated poles of equal strength can be constructed by using the expansion of cotangent into partial fractions \cite{Markushevich2005}
\begin{equation}
\label{eqTransmCot}
T(\omega) =
\sum\limits_{p = 1}^P {
\frac{\pi \sigma_p} {4 \omega_{\rm{qw}}}
\cot \left[ \frac{\pi (\omega - \omega_p)} {4 \omega_{\rm{qw}}} \right].
}
\end{equation}
Alternatively, one can operate with multiple zeros rather than poles and to rewrite the transmission as a product
\begin{equation}
\label{eqTransmSin}
T(\omega) =
\prod\limits_{p = 1}^P {
\frac
{\sin [\pi \omega_p /(4\omega_{\rm{qw}})]}
{\sin [\pi (\omega - \omega_p)/(4 \omega_{\rm{qw}})]}
}.
\end{equation}
The formulas (\ref{eqTransmExp})--({\ref{eqTransmSin}), are fully equivalent to each other. For example, to derive Eq.~(\ref{eqTransmCot}) from Eq.~(\ref{eqTransmExp}), it is sufficient to notice that due to causality $a_0  = \sum\nolimits_{p = 1}^P {\sigma_p} = 0$.

To analyze the propagation of ultrashort pulses through the structure, it is important to compute the group delay as a function of frequency \cite{Zhukovsky2008}. By separating the amplitude and phase in the transmission spectrum $T(\omega ) = |T(\omega)| \exp (i \varphi )$, the group delay can be defined as $\tau_{\rm{D}} (\omega) = {\rm{d}} \varphi / {\rm{d}} \omega  = {\rm{Im}} [T'(\omega) / T(\omega)]$. An explicit formula for it can be obtained particularly easy from Eq.~(\ref{eqTransmSin})
\begin{equation}
\label{eqGroupDelay}
\tau_{\rm{D}} (\omega) =
- \frac {\pi} {4 \omega_{\rm{qw}}}
{\rm{Im}}
\left(
\sum\limits_{p = 1}^P {\cot
\left[ \frac {\pi (\omega - \omega_p)} {4 \omega_{\rm{qw}}} \right]
}
\right).
\end{equation}
Since the integral $\int_{-\pi/2}^{\pi/2} {\cot(z - z_p) {\rm{d}} z} = - i \pi$ for any $z_p$ in the lower part of the complex plane ${\rm{Im}}[z_p] < 0$, the averaged variations of the group delay do not depend on the particular arrangement of the layers in the structure $
\int_{- 2 \omega_{\rm{qw}}}^{2 \omega_{\rm{qw}}}
{\tau_{\rm{D}} (\omega) {\rm{d}} \omega} = \pi P
$. On the other hand, for a homogeneous slab consisting of $M$ quarter-wave layers $
\int_{- 2 \omega_{\rm{qw}}}^{2 \omega_{\rm{qw}}}
{({\rm{d}} \varphi / {\rm{d}} \omega ){\rm{d}} \omega } = 2 \pi M
$. Therefore, $P = 2M$, or the number of resonances in the interval $(-2 \omega_{\rm{qw}}, 2 \omega_{\rm{qw}}]$ equals twice the number of layers [cf.~Fig.~\ref{fig1}].

The concept of group delay is closely related to other characteristics of multilayered structures such as the group velocity or the density of modes. This means that similar analytical formulas can be derived for them. For instance, the group velocity (or traversal velocity) can be defined as $v_{\rm{gr}} = L / \tau_{\rm{D}}$, where $L$ is the total length of the structure. It can be also rewritten as $v_{\rm{gr}} / c = \!M \tau / (\tau_{\rm{D}} \langle n \rangle)$, where an averaged refractive index $\langle n \rangle = (\sum\nolimits_{m = 1}^M {n_m d_m }) / L$ was introduced. This makes easier the comparison of the data for the group delay shown in Fig.~\ref{fig1} with the group velocity.

In general, the transmitted signal $f_{\rm{out}}(t)$ can be found as a convolution of $G(t)$ with the input signal $f_{\rm{in}}(t)$:
\begin{equation}
\label{eqConvolution}
f_{\rm{out}}(t) \!=\!\!\! \int_{-\infty}^{+\infty} {\!\!\!\!\!G(t - t')f_{\rm{in}} (t') {\rm{d}} t'}
\!=\! \sum\limits_{n = 0}^\infty {a_n f_{\rm{in}} (t - n \tau)}.\!
\end{equation}
This series converges very rapidly because the coefficients $a_n$ decay exponentially with increasing $n$. The simplest reshaping of the signals takes place when only one resonance is excited. As an example, we consider a well isolated resonance at $\omega_p \approx \omega_{\rm{qw}}$ that exists for the structure shown in Fig.~\ref{fig1}(d). It is worth noting that the group delay experienced by the pulses with the central frequency close to ${\rm Re}[\omega_p]$ can be estimated as $
\tau_{\rm{D}}({\rm Re} [\omega_p ]) / \tau
\approx
-2 / (\pi {\rm Im} [\omega_p / \omega_{\rm{qw}}])
$, and it coincides with the decay time of the resonance in the absence of excitation. Which effect will dominate depends on the duration of the incident pulse [see Fig.~\ref{fig2}]. The exponential stretching is more pronounced for shorter pulses, and it can be accompanied with beating if several resonances are located close to each other~\cite{DalNegro2003}.

\begin{figure}[t]
\centerline{\includegraphics[width=80mm]{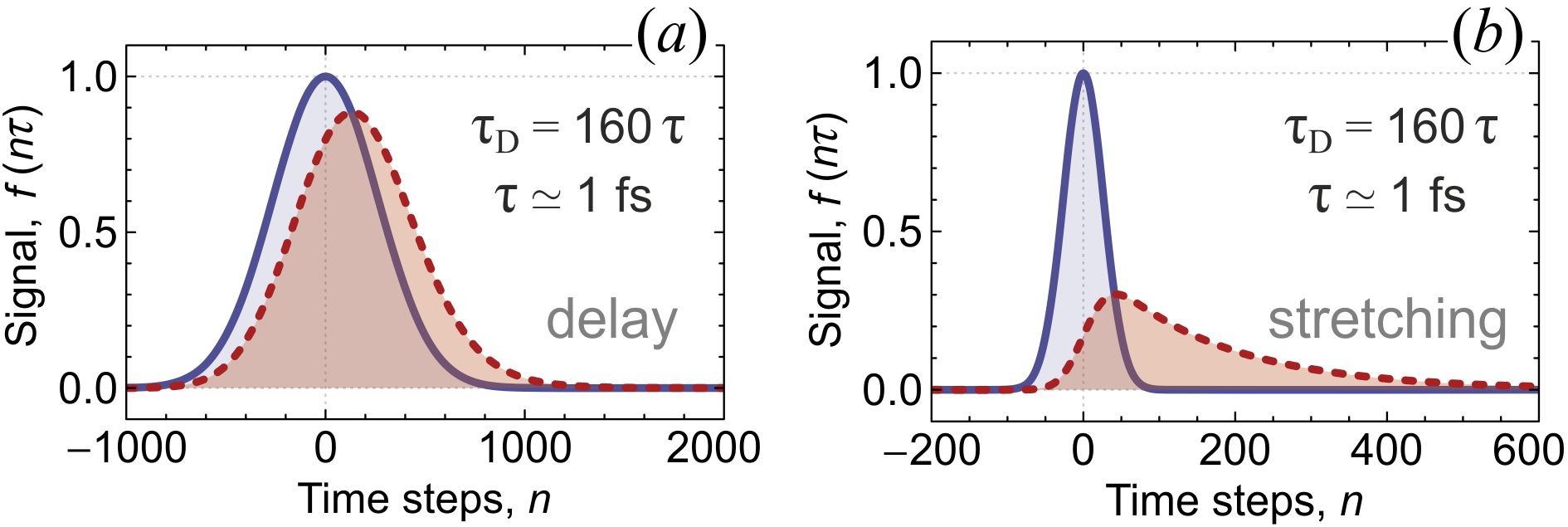}}
\caption{\small{\label{fig2} (Color online) Gaussian pulses before (solid) and after (dashed) propagation through a multilayered structure.
}}
\end{figure}

\begin{figure}[b]
\centerline{\includegraphics[width=80mm]{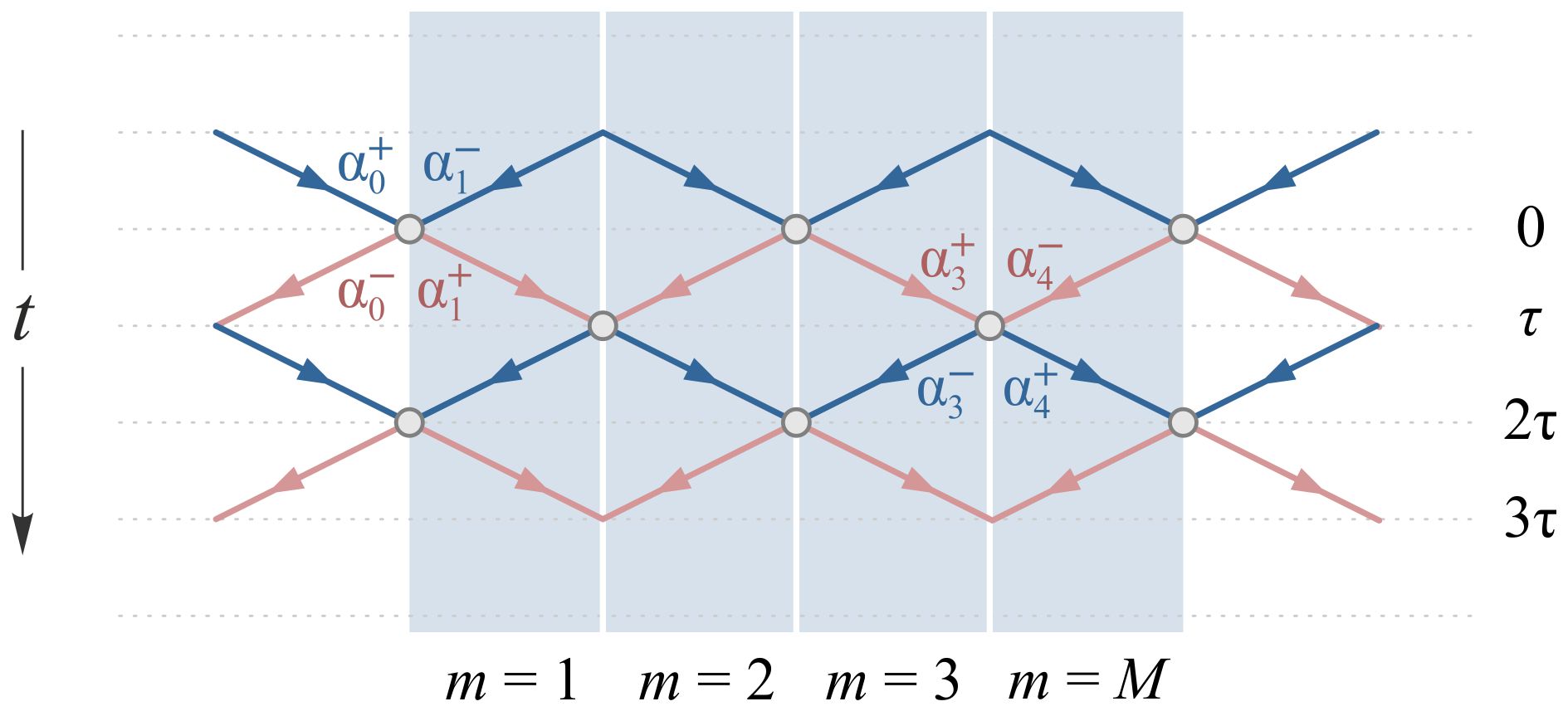}}
\caption{\small{\label{fig3} (Color online)
Propagation of rays inside a quarter-wave multilayered structure.
}}
\end{figure}

Although the positions of the poles can be found by iterations using the transfer matrix method, we developed a more reliable and efficient algorithm which is based on considering the ray propagation inside the structure.

Initially, the amplitudes of forward and backward moving rays in all layers of the structure and on the boundaries are set to zero $\alpha_m^\pm = 0$ for $0 \le m \le M + 1$ [see Fig.~\ref{fig3}]. Then, a short probe signal is launched from the left, which corresponds to setting $\alpha_0^+  = 1$ only for the first moment of time $t = 0$. Afterwards, a time marching scheme is applied to collect the outgoing signals on the right boundary at discrete moments of time $t = n \tau$. For odd (even) time steps $n$, one should apply $
[\alpha_m^-, \; \alpha_{m + 1}^+]^{\rm{T}} = {\bf{Y}}_{m,m + 1}
[\alpha_m^+, \; \alpha_{m + 1}^-]^{\rm{T}}
$ with index $m$ running over even (odd) numbers. The matrix ${\bf{Y}}_{u,v}$ consists of the Fresnel coefficients which describe the scattering of rays at the interface between media with the refractive indices $n_u$ and $n_v$
\begin{equation}
\label{eqY}
{\bf{Y}}_{u,v} =
\frac{1} {n_u + n_v}
\left[ \begin{array}{cc}
   {n_u - n_v}  &  {2 n_v}  \\
   {2 n_u}  &  {n_v - n_u}  \\
\end{array} \right].
\end{equation}

Since the output signal on the right can be represented as a sum of exponentially decaying modes $[{\bf{A}}]_n  = - \sum\nolimits_{p = 1}^P {i \sigma_p \tau \exp (- i n \omega_p \tau)}$, the unknown amplitudes $\sigma_p$ and resonant frequencies $\omega_p$ can be found as a solution of the following eigenvalue problem \cite{Mandelshtam2001}
\begin{equation}
\label{eqEV}
\hat{\bf{H}}_1 {\bf{V}}_p
= \exp (- i \omega_p \tau) \hat{\bf{H}}_0 {\bf{V}}_p,
\end{equation}
where the matrix elements of auxiliary Hamiltonians ${\hat{\bf{H}}}_k$ are defined as $[{\hat{\bf{H}}}_k]_{u,v} = [{\bf{A}}]_{u + v + k}$. This harmonic inversion technique is very efficient because it requires the knowledge of only $2P$ elements in the vector $\bf{A}$. The poles $\omega_p$ are directly related to the eigenvalues of Eq.~(\ref{eqEV}), and their strength $\sigma_p$ can be determined after a proper normalization of the eigenvectors ${\bf{V}}_p$
\begin{equation}
\label{eqNorm}
\sigma_p = \frac {i} {\tau}
\,
\frac
{[({\bf{V}}_p )^{\rm{T}} \cdot {\bf{A}}]^2 }
{ ({\bf{V}}_p )^{\rm{T}} \cdot (\hat{\bf{H}}_0 {\bf{V}}_p )}.
\end{equation}

In conclusion, we showed that the transmission spectrum of multilayered structures can be described by analytical formulas regardless of the specific arrangement of the layers. These formulas use the position of resonances as parameters, and we presented an efficient algorithm to find them. This establishes a new approach to analyze the transmission properties of multilayers. It can be useful not only for an advanced reshaping of ultrashort pulses but also for the improvement of nonlinear effects such as self-pulsing or nonreciprocal transmission \cite{Grigoriev2011a,Grigoriev2011}.

%


\end{document}